\documentclass[journal,transsp,10pt]{IEEEtran}
\usepackage[dvips]{graphicx}
\usepackage{color,soul,amsmath,amssymb}
\usepackage[usenames,dvipsnames]{xcolor}
\usepackage{algorithmic}
\usepackage{algorithm}
\usepackage[utf8]{inputenc}
\usepackage[english]{babel}
\usepackage{amsthm}

\newtheorem{theorem}{Theorem}

\newcommand{\am}{\mathbf{a}}
\newcommand{\e}{\mathbf{e}}

\newcommand{\x}{\mathbf{x}}

\newcommand{\xh}{\widehat{\mathbf{x}}}

\newcommand{\y}{\mathbf{y}}

\newcommand{\A}{\mathbf{A}}

\newcommand{\ZZ}{\mathbf{Z}}

\newcommand{\Z}{\mathcal{Z}}
\newcommand{\RR}{\Bbb{R}}

\newcommand{\R}{\mathbb{R}}

\newcommand{\rr}{\mathbf{r}}
\newcommand{\argmin}[1]{\underset{#1}{\operatorname{arg\,min}}}

\newcommand{\xheps}{\xh^{\tau_R}}
\newcommand{\xho}{\xh^{\tau_R}}

\providecommand{\norm}[1]{\lVert#1\rVert}

\begin{document}

\title{The regularized tau estimator: A robust and efficient solution to ill-posed linear inverse problems}

\author{\IEEEauthorblockN{Marta Martinez-Camara\IEEEauthorrefmark{1},
Michael Muma\IEEEauthorrefmark{2},
Benjam\'in B\'ejar\IEEEauthorrefmark{1},
Abdelhak M. Zoubir\IEEEauthorrefmark{2}, 
and Martin Vetterli\IEEEauthorrefmark{1} \vspace{10 pt}}\\


}


\IEEEtitleabstractindextext{%
\begin{abstract}
Linear inverse problems are ubiquitous. Often the measurements do not
follow a Gaussian distribution. Additionally, a model matrix with a
large condition number can complicate the problem further by making it
ill-posed. In this case, the performance of popular estimators may
deteriorate significantly.  We have developed a new estimator that is
both nearly optimal in the presence of Gaussian errors while being also
robust against outliers. Furthermore, it obtains meaningful estimates
when the problem is ill-posed through the inclusion of $\ell_1$ and
$\ell_2$ regularizations. The computation of our estimate involves
minimizing a non-convex objective function. Hence, we are not guaranteed
to find the global minimum in a reasonable amount of time. Thus, we
propose two algorithms that converge to a good local minimum in a
reasonable (and adjustable) amount of time, as an approximation of the
global minimum. We also analyze how the introduction of the
regularization term affects the statistical properties of our estimator.
We confirm high robustness against outliers and asymptotic efficiency for
Gaussian distributions by deriving measures of robustness such as the
influence function, sensitivity curve, bias, asymptotic variance, and
mean square error. We verify the theoretical results using numerical
experiments and show that the proposed estimator outperforms recently
proposed methods, especially for increasing amounts of outlier
contamination. Python code for all of the algorithms are
available online in the spirit of reproducible research.
\end{abstract}

}

\maketitle

\IEEEdisplaynontitleabstractindextext

%
\IEEEpeerreviewmaketitle

\section{Introduction}

Linear inverse problems are ubiquitous, but in spite of their simple formulation, they have kept researchers busy for decades. Scarce and noisy measurements or ill-posedness substantially complicate their solution.  

In a linear inverse problem, we wish to find a vector $\x_0\in\RR^{n\times 1}$ from a set of measurements $\y$, given as 
\begin{equation}
\y = \A\x_0 + \e,
\label{eq:signalmodel}
\end{equation}
where we call $\y\in\RR^{m\times 1}$ the measurements or data, $\A \in \RR^{m\times n}$ is the  the model matrix, and $\e\in\RR^{m\times 1}$ is an additive error term. The measurements $\y$ are known and the model matrix $\A$ is usually known or can be estimated; the errors $\e$ and the source $\x_0$ are unknown. 

The common approach to estimate $\x_0$ is to use the least square (LS) estimator. This estimator finds the $\x$ that minimizes the $\ell_2$ norm of the residuals $\rr(\x)= \y - \A\x$, i.e. 
\begin{align}
\xh_{LS} = \argmin{\x}\; \norm{\y - \A\x}_2^2.
\end{align}

But the LS estimator does not always work as desired. Two difficulties may arise: ill-posedness and outliers.

First, if the model matrix $\A$ has a large condition number, $\xh_{LS}$ is very sensitive to the error $\e$. Then, even an $\e$ with a small norm can produce a large deviation of $\xh_{LS}$ from the ground truth $\x_0$. The problem is then said to be ill-posed \cite{ribes08}.

The LS estimator is the maximum likelihood estimator if the errors $\e$ are sampled from a population with a Gaussian distribution, which is not always the case. Often, the components of $\e$ come from a heavy-tailed distribution, i.e. they contain outliers. These large components of $\e$ can cause $\xh_{LS}$ to deviate strongly from the true value $\x_0$.

\begin{figure}[ht!]
\centering
  \includegraphics[width=1\linewidth]{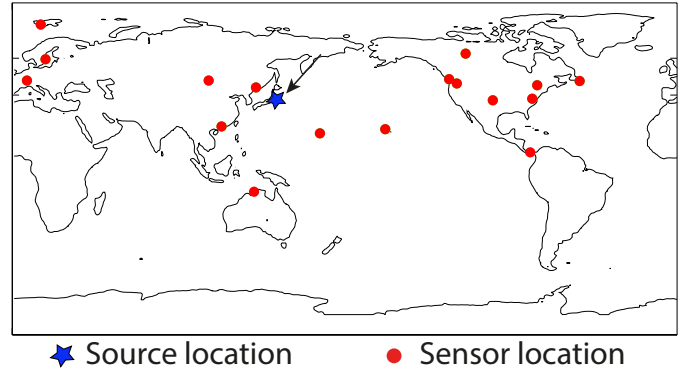}
  \caption{Example of estimation of releases to the atmosphere problem. The blue dot represents the spatial location of the source of the releases. The red dots are the locations of the sensors. The goal is to estimate the temporal variations of the source using concentration measurements collected by the sensors over time.}
   \vspace{-0.5em}
  \label{fig:emissions}
\end{figure}

One application where these two difficulties may appear simultaneously is in the estimation of the temporal releases of a pollutant to the atmosphere using temporal measurements of the concentration of the pollutant in the air taken at different locations (see Figure \ref{fig:emissions}).

This situation can be formulated as a linear inverse problem where $\y$ contains the measurements that we collect, $\A$ can be estimated using atmospheric dispersion models and meteorological information, and $\x_0$ describes the temporal emissions at a chosen spatial point. Unfortunately, the sparsity of sensors and unfavourable weather patterns can cause the matrix $\A$ to have a large condition number. At the same time, errors in the sensors and in the model can provoke large differences between model and reality, which in turn may cause the errors $\e$ to be heavy-tailed and to contain outliers \cite{marta14}. 

Robust estimators like the M or S  estimators \cite{maronna06} have a smaller bias and variance than the LS estimator when outliers are present in the data. The drawback is that, in general, when they are tuned for robustness against ourliers, their variance is larger than that of the LS estimator when there are no outliers in the data, i.e., the distribution of the errors is Gaussian. The $\tau$ estimator \cite{yohai88} improves this trade off: by adapting automatically to the distribution of the data, it is robust against outliers while having a variance close to the LS estimator when the errors have a Gaussian distribution. This also holds for the MM estimator \cite{yohai1987mm} which combines an S estimation step with an M estimation step.

However, the M, S, MM or $\tau$ estimators are ill-suited for ill-posed problems. That is why regularized robust estimators, designed to cope with linear inverse problems that are ill-posed and contain outliers, have been proposed \cite{tharmaratnam10,alfons2013sparselts,kimkoivunen2015rankslasso,loh2015regM,croux15,smucler2015regMM,hoffmann2015sparseparM}. First results on asymptotic and robustness theory for the M estimator \cite{croux15,loh2015regM}, the S estimator \cite{smucler2015regMM}, and the MM estimator \cite{smucler2015regMM} have been obtained very recently. The mean-squared error (MSE) of these estimators in the presence of outliers is smaller than that of regularized estimators that use the LS loss function.

In this paper, we propose a new regularized robust estimator: the $\tau$ estimator. We study how the statistical properties of the $\tau$ estimator are affected by different regularizations, and we compare its performance with recently proposed robust regularized estimators. We also give algorithms to compute these estimates, and we provide an analysis of their performance using simulated data.

%
%
%
%

This paper is organized as follows: In Section \ref{sec:proposed_estimator}, we propose our new robust regularized estimator, and we explain its underlying intuition. We also propose different heuristic algorithms to compute the new estimates. In Section \ref{sec:analysis}, we give an analysis of the robustness and efficiency of the new estimator. Derivations and proofs are provided in the Appendices A-C. Finally, Section \ref{sec:conclusion} concludes the paper.

\section {Proposed estimator}
\label{sec:proposed_estimator}
\subsection{The $\tau$ estimator}

The $\tau$ estimator \cite{yohai88} is simultaneously robust to outliers and efficient w.r.t. the Gaussian distribution. The efficiency is defined as the asymptotic variance of the maximum likelihood estimator for the data model divided by the asymptotic variance of the estimator under consideration \cite{maronna06}. It takes values between 0 and 1, where 1 is the highest possible efficiency. The $\tau$ estimator can handle up to 50\% of outliers in the data (achieving a breakdown point of 0.5) and at the same time it performs almost as well as the LS estimator when the errors are Gaussian. In other words, its asymptotic efficiency at the normal distribution is close to one.

\begin{figure}[ht!]
\centering
  \includegraphics[width=0.9\linewidth]{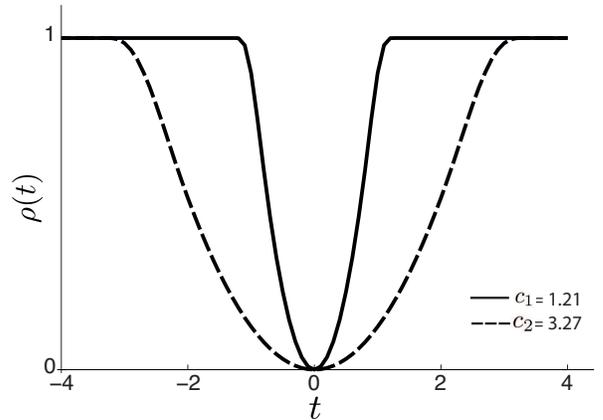}
  \caption{Example of optimal $\rho$-functions that can be used in the $\tau$ estimator.}
  \label{fig:rhosopt}
  \vspace{-0.8em}
\end{figure}

To understand the intuition behind the $\tau$ estimator, we first briefly revisit the M estimator \cite{maronna06}, which is defined as
\begin{align}
\xh_m^M = \argmin{\x}\;\frac{1}{m}\sum_{i=1}^m \rho \left(\frac{r_i(\x)}{{\hat{\sigma}}}\right),
\label{eq:mestimator}
\end{align}
where $r_i(\x)$ is the $i$-th component of $\rr(\x)$ and ${\hat{\sigma}}$ is an estimate of the scale of the errors $\e_i$. $\rho(\cdot)$ is a function such that 
\begin{enumerate}                                                                                                                                                     
\item $\rho(x)$ is a nondecreasing function of $|x|$,
\item $\rho(0)=0$
\item $\rho(x)$ increasing for $x>0$ such that $\rho(x)< \rho(\infty)$
\item If $\rho$ is bounded, it is assumed that $\rho(\infty)=1$.
\end{enumerate}
 We can see two examples of this function in Figure \ref{fig:rhosopt}. The function drawn as a solid line produces a more robust, but less efficient, M estimator than the one using the function drawn as a dashed line. 

The $\tau$ estimator has been shown in \cite{yohai88} to be equivalent to an M estimator whose $\rho$ function is the weighted sum of two other $\rho$ functions
\begin{equation}
\rho_\tau(u) = w_m (\rr(\x))\rho_1(u) + \rho_2(u)
\label{eq:rhotau}
\end{equation}
for some weight function $w_m(\cdot)$. 
The interesting thing is that the non-negative weights $w_m (\x)$ (that we will define later) adapt automatically to the distribution of the data. Then, if we choose $\rho_1$ to be a robust loss function, and $\rho_2$ to be an efficient one, the $\tau$ estimator will have a combination of properties, depending on the distribution of the data: if there are no outliers, the weights $w_m (\rr(\x))$ will be approximately zero and the estimator will be efficient; if there are many outliers, $w_m (\rr(\x))$ will be large and the estimator will be robust. 

Although the $\tau$ estimate of regression is equivalent to an M estimate, it is defined as the minimizer of a particular robust and efficient estimate of the scale of the residuals, the $\tau$-scale estimate $\hat{\sigma}_\tau$  \cite{yohai88}
\begin{align}
\xh_m^\tau = &\argmin{\x}\;\hat{\sigma}^2_\tau(\rr(\x)),
\label{eq:regtau}
\end{align}
where the $\tau$ scale estimate is defined as
\begin{equation}
\hat{\sigma}_\tau^2(\rr(\x)) = \hat{\sigma}_M^2 (\rr(\x))\frac{1}{m}\sum_{i=1}^m \rho_2\left(\frac{r_i(\x)}{\hat{\sigma}_M(\rr(\x))}\right).
\label{eq:tausigma}
\end{equation}
Here, $\hat{\sigma}_M (\rr(\x))$ is another robust scale estimate, the M scale estimate \cite{maronna06}, that satisfies
\begin{equation}
\frac{1}{m}\sum_{i=1}^m \rho_1\left(\frac{r_i(\x)}{\hat{\sigma}_M(\rr(\x))}\right) = b, 
\label{eq:robscale}
\end{equation}
with $b = E_{H_0}[\rho_1(u)]$, such that $E_{H_0}[\cdot]$ denotes expectation w.r.t. the standard Gaussian distribution $H_0$.

However, in spite of all its good properties, the $\tau$ estimator cannot deal with ill-posed problems \cite{marta15}. 
\subsection{On ill-possedness and regularization}

Hadamard defined a well-posed problem as one whose solution exists, is unique and changes continuously with the initial conditions \cite{hadamard02}. If any of these conditions is violated, the problem is ill-posed. 

In the linear inverse problem that we are studying, the third condition is violated: when the condition number of $\A$ is too large, the LS, M, MM, or $\tau$ estimates are too sensitive to any small deviation in the measurements.

The way to transform an ill-posed problem into a well-posed one is to include more \emph{a priori} information on the solution $\x_0$ into the problem, that is to \emph{regularize} the problem.

There are different possible regularizations for a problem. One typical choice is the Tikhonov regularization, which looks for solutions with low energy, i.e., with a small $\ell_2$ norm. It achieves this by adding a penalty $\norm{\x}_2$ to the loss function being minimized.
 

In the past 20 years, another regularization has been used frequently, namely the sparse regularization \cite{elad09}. It looks for sparse solutions, i.e. solutions with just a few non-zero components. Since the $\ell_0$ norm counts the number of non-zero components in a vector, a sparse regularization looks for solutions with a small $\ell_0$ norm.

This norm being non-convex, its minimizing is an NP hard problem. To make the computation tractable, we can relax the problem by replacing the $\ell_0$ norm by its closest convex norm, the $\ell_1$ norm. Under certain conditions~\cite{elad09}, the solution of the original problem and the solution of the relaxed one are the same.


\subsection{The regularized $\tau$ estimator}
Our purpose is to generalize the $\tau$ estimator to make it suitable for ill-posed linear inverse problems with outliers. For that, we add a regularization term to the $\tau$ scale loss function

\begin{eqnarray}
\xh_m^{\tau_R} &=&  \argmin{\x}\;\tau_R^2(\rr(\x))\nonumber \\ 
	    &=&   \argmin{\x}\;\hat{\sigma}^2_\tau(\rr(\x))\! + \! \lambda\sum_{i=1}^n J(x_i),
\label{eq:regtau}
\end{eqnarray} 
where $x_i$ is the $i$-th component of $\x$, which has $n$ components.
We will focus on two regularizations: the Tikhonov regularization, that uses a differentiable $J(x) = x^2$, and the sparse regularization that uses a non differentiable $J(x) = |x|$.

\section{Algorithms}
Before studying the properties of our proposed estimator, we need to know how to compute it. For that, we have to minimize the objective function (\ref{eq:regtau}). This function is non-convex, so we do not have any guarantee of finding the global minimum in a finite amount of time. Hence, we propose heuristic algorithms that compute an approximation of this global minimum in a reasonable amount of time.

We will explain the algorithms in three parts: how to find local minima, how to approximate the global minimum, and how to reduce the computational cost.

\subsection{Finding local minima}
The derivative of the objective function (\ref{eq:regtau}) is equal to zero at all local minima. So, the local minima are given as the solutions of the equation
\begin{equation}
\frac{\partial \tau_R^2(\rr(\x))}{\partial \x} = \frac{\partial (\hat{\sigma}_\tau^2(\rr(\x)) + \lambda \sum_{i=1}^n J(x_i))}{\partial \x} = 0.
\label{eq:tauderivative}
\end{equation}

In Appendix \ref{ap:localminima}, we show that if the regularization function $J(x)$ is a first order differentiable function, then the derivative of the regularized $\tau$ objective function (\ref{eq:regtau}) is also the derivative of a penalized iterative reweighted least squares algorithm (IRLS) that minimizes the objective function  

\begin{equation} 
\xh_m^{\tau_R} = \argmin{\x}\;\norm{\ZZ^{1/2}(\x)(\A\x - \y)}^2_2 + \lambda \sum_{i=1}^n J(x_i),
\label{eq:WLS}
\end{equation}
where $\ZZ(\x)\in\RR^{m\times n}$ is a rectangular diagonal matrix with diagonal components 
\begin{align}
z_i = \begin{cases} \psi_{\tau}(\tilde{r}_i(\x)) / 2\tilde{r}_i(\x) & \quad \text{if } \tilde{r}_i(\x)\neq0 \\ 
0 & \quad \text{otherwise. } \\ \end{cases} 
\label{eq:weights}
\end{align}
Here
\begin{align}
 \tilde{r}_i(\x) &:= \frac{r_i(\x)}{\hat{\sigma}_M(\rr(\x))}, \\
 \psi_\tau (u)&:= \frac{\partial \rho_\tau(u)}{\partial u},
\label{eq:psi}
\end{align}
where $\hat{\sigma}_M(u)$ satisfies (\ref{eq:robscale}), and $\rho_\tau(u)$ is as defined in (\ref{eq:rhotau}), using
\begingroup\makeatletter\def\f@size{9}\check@mathfonts
\begin{align}
w_m(\x) =  \frac{\sum_{i=1}^m -2\rho_2\left(\tilde{r}_i(\x)\right)  + \psi_2\left(\tilde{r}_i(\x)\right)\tilde{r}_i(\x)}{\sum_{i=1}^m
\psi_1\left(\tilde{r}_i(\x)\right) \tilde{r}_i(\x)}
\end{align}
\endgroup
Hence, this penalized IRLS has the same minima as (\ref{eq:regtau}), and we use it to find such minima.  Details of this penalized IRLS algorithm are provided in Algorithm \ref{alg:irls}.
\begin{algorithm}[H]
\caption{Regularized IRLS}
\label{alg:irls}
\begin{algorithmic}
\vspace{5 pt}
\STATE INPUT: $\y\in\mathbb{R}^{m}$, $\A\in\mathbb{R}^{m\times n}$, $\lambda$, $\xi$, $\x[0]$, $K$
\STATE OUTPUT: $\xh$
\STATE \textbf{constrain} $\lambda \geq 0$, $\xi\geq 0$, $K\geq 0$
\FOR{$k=0\;$\TO $K-1$}
\STATE $\rr[k]\gets\y -\A\x[k]$
\STATE Compute $\hat{\sigma}_M[k]$
\STATE Compute $\ZZ[k]$
\STATE $\x[k+1]\gets(\A^\top\ZZ[k]\A+\lambda^2)^{-1}\A^\top\ZZ[k]\y$
\IF{$||\x[k+1] - \x[k]||< \xi$}
\STATE \textbf{break}
\ENDIF
\ENDFOR
\RETURN $\xh\gets\x[k+1]$
\vspace{5 pt}
\end{algorithmic}
\end{algorithm}

But when the $\tau$ estimator is regularized with the $\ell_1$ norm, i.e. $J(x) = |x|$, $J(x)$ is not differentiable, and we cannot directly apply the equivalence that we explained above. Nevertheless, by approximating $|x|$ with a differentiable function, we show in Appendix \ref{ap:localminima} that a local minimum in (\ref{eq:regtau}) with $J(x) = |x|$ is indeed also a local minimum in (\ref{eq:WLS}) with $J(x) = |x|$. So, we can use a regularized IRLS algorithm to find the local minima in this case as well.

\subsection{Approximating the global minimum}

\begin{figure}[ht!]
\centering
  \includegraphics[width=0.9\linewidth]{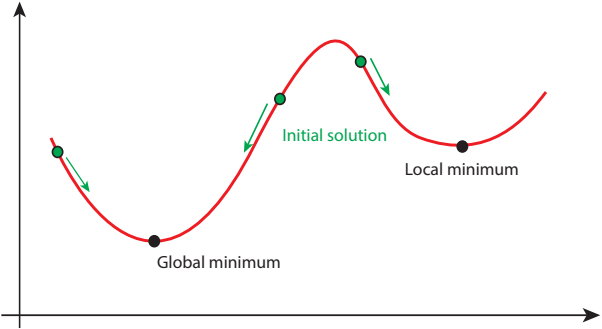}
  \caption{Example of a non convex function: the local minima that we find with the IRLS algorithm depends on the initial solution that we use.}
  \label{fig:nonconvex}
  \vspace{-0.8em}
\end{figure}

As the objective function (\ref{eq:regtau}) is non-convex, each local minimum that we find using the IRLS algorithm depends on the initial solution that we use (see Figure \ref{fig:nonconvex}). This takes us to the second step of the algorithm: to approximate the global minimum. For that, we run the IRLS algorithm $Q$ times using for each run a different random initial solution. Then, we compare the different minima that we found and pick the best one. See Algorithm \ref{alg:basic} for more details.

\begin{algorithm}[H]
\caption{Basic algorithm}
\label{alg:basic}
\begin{algorithmic}
\vspace{5 pt}
\STATE INPUT: $\y\in\mathbb{R}^{m}$, $\A\in\mathbb{R}^{m\times n}$, $\lambda$, $\xi$, $K$, $Q$
\STATE OUTPUT: $\xh$
\STATE \textbf{constrain} $\lambda \geq 0$, $\xi\geq 0$, $K\geq 0$, $Q\geq 0$
\FOR{$q=0\;$\TO $Q-1$}
\STATE $\x_q[0]\gets$random initial condition
\STATE $\xh_q\gets$Algorithm~\ref{alg:irls} ($\y,\A,\lambda,\xi,\x_q[0],K$)
\ENDFOR
\RETURN $\xh\gets\argmin{\xh_q}\;\tau_R^2 (\rr(\xh_q))$
\vspace{5 pt}
\end{algorithmic}
\end{algorithm}

\subsection{Saving computational cost}
Approximation of the global minimum is computationally expensive because the IRLS algorithm needs to be run many times (i.e. $Q$ is large). This takes us to the third step: saving computational cost. For that, we do the following: for each initial solution, we run just a few iterations of the IRLS algorithm; the convergence at the beginning is fast. We keep the $K$ best minima. In the second phase, we use these $K$ minima as initial solutions. This time, we run the IRLS algorithm until it reaches convergence. Finally, we select the best minimum that we found. See Algorithm \ref{alg:fast} for more details. This is one of the simplest algorithms that we can use to find a global minimum when the objective function is not convex. But, as we will see in the next sections, it works well in the tested scenarios.
\begin{algorithm}[H]
\caption{Fast algorithm}
\label{alg:fast}
\begin{algorithmic}
\vspace{5 pt}
\STATE INPUT: $\y\in\mathbb{R}^{m}$, $\A\in\mathbb{R}^{m\times n}$, $\lambda$, $\xi$, $K$, $Q$,$M$
\STATE OUTPUT: $\xh$
\STATE \textbf{constrain} $\lambda \geq 0$, $\xi\geq 0$, $K\geq 0$, $Q\geq 0$, $M\geq 0$
\FOR{$q=0\;$\TO $Q-1$}
\STATE $\x_q[0]\gets$random initial condition
\STATE $\xh_q\gets$Algorithm~\ref{alg:irls} ($\y,\A,\lambda,\xi,\x_q[0],K$)
\ENDFOR
\STATE $\Z\gets$set of $M$ best solutions $\xh_q$ 
\FOR{$\x_z\in\Z$}
\STATE $\xh_z\gets$Algorithm~\ref{alg:irls} ($\y,\A,\lambda,\xi,\x_z,\infty$)
\ENDFOR
\RETURN $\xh\gets\argmin{\xh_z}\;\tau_R^2 (\rr(\xh_z))$
\vspace{5 pt}
\end{algorithmic}
\end{algorithm}

\section{Analysis}
\label{sec:analysis}
In this section we study the properties of the proposed regularized $\tau$ estimator. In particular, we study two aspects: the robustness of the estimator against outliers and the performance of the estimator at the nominal distribution. For the first task, we derive the influence function (IF) of the estimator, we compare it with its sensitivity curve (SC), and we explore the break down point of the estimator. For the second task, we study the asymptotic variance (ASV) and the bias of the estimator when the errors have a Gaussian distribution.


In this analysis, we work in the asymptotic regime, where we assume the number of measurements $m$ to tend to infinity, and the number of unknowns $n$ remains fixed and finite. Let the measurements $y$ and the errors $e$ be modelled by random variables
\begin{align} 
y = \am\x_0 + e,
\label{eq:basicModel}
\end{align}
where $\am\in\RR^{1\times n}$ is a row vector of i.i.d. random entries independent of $e$, and $\x_0\in\RR^{n\times 1}$ is the deterministic vector of unknown true parameters. We assume $\am$ and $y$ to have a joint distribution that we denote by $H$. When the measurements and/or the model contain outliers, $H$ takes the form 
\begin{equation}
H_\epsilon = (1-\epsilon)H_{0} + \epsilon G,
\end{equation}
where $H_{0}$ is the distribution of the \emph{clean} data, $G$ is any distribution different from $H_{0}$, and $\epsilon$ is the proportion of outliers in the data.

In this asymptotic regime, the M-scale estimate $\hat{\sigma}_M(u)$ is defined by
 \begin{equation}
\mathbb{E}_H\left[ \rho_1\left(\frac{u}{\hat{\sigma}_M(u)}\right)\right] = b, 
\label{eq:robscaleAs1}
\end{equation}
where $b = E_{H_0}[\rho_1(u)]$, and the regularized $\tau$ estimate becomes 
\begin{equation}
\xh^{\tau_R} = \argmin{\x}\;\hat{\sigma}_M^2 (r(\x)) \mathbb{E}_H \left[\rho_2\left(\frac{r(\x)}{\hat{\sigma}_M(r(\x))}\right)\right]+ \lambda J(\x),
\label{eq:asTau}
\end{equation}
where $r(\x) = y - \am\x$.
%
%
%
%

\subsection{A few outliers}

The IF is a measure of robustness \cite{zoubir12}. It was originally proposed by F. R. Hampel \cite{hampel74}. It indicates how an estimate changes when there is an infinitesimal proportion of outliers $\epsilon$ in the data. The IF is defined as the derivative of the asymptotic estimate $\xh$ w.r.t. $\epsilon$ when 
\begin{enumerate}
\item $H_\epsilon = (1-\epsilon)H_0 + \epsilon G$
\item $H_0$ is any nominal distribution.
\item $G = \delta_{\am_0,y_0}$ is the point mass at $(\am_0,y_0)$ 
\item Particularized to $\epsilon$ equals zero.
\end{enumerate}
More formally we can write
\begin{equation}
\text{IF}((\am_0,y_0), \xh, H_0) = \left.\frac{\partial[\xh((1-\epsilon)H_0 + \epsilon\delta_{\am_0,y_0})]}{\partial \epsilon}\right|_{\epsilon = 0}.
\label{eq:if}
\end{equation}

The IF helps us understand what happens when we add one more observation with value $(\am_0, y_0)$ to a very large sample. If $\epsilon$ is small, the asymptotic bias of the estimator that is caused by adding such observation $(\am_0, y_0)$ can be approximated by $\epsilon\text{IF}((\am_0,y_0), \xh, H_0)$ \cite{maronna06}. Desirable properties of the IF of an estimator are boundedness and continuity.

\begin{theorem}
 Let $y = \am\x_0 + e$ be as given in (\ref{eq:basicModel}). Let also $J,\rho_1,\rho_2$ be twice differentiable functions. Assume the M-scale estimate $\hat{\sigma}_M(u)$ as given in (\ref{eq:robscaleAs1}). Define $\tilde{r}(\x):=r(\x)/\hat{\sigma}_M(r(\x)) $ and $\psi_i(u) = \rho'_i(u)$. Then the influence function of the regularized $\tau$ estimator $\xh^{\tau_R}$ is given by 
\begin{align}
&\text{IF}((\am_0,y_0), \xh^{\tau_R}, H_0) = \nonumber\\ 
&\left(-\mathbb{E}_{H_0} \left[\am^\top\frac{\partial (\psi_\tau(\tilde{r}(\xho))\hat{\sigma}_M(r(\xho)))}{\partial \xho}\right]\right. \nonumber\\ 
&+\lambda J''(\xho)\bigg)^{-1}\times \nonumber\\ 
& \bigg(\psi_\tau\left(\frac{y_0 - \am_0\xho}{\hat{\sigma}_M(r(\xho))}\right)\hat{\sigma}_M(r(\xho))\am_0^\top \nonumber\\
&\mathbb{-E}_{H_0}\left[\left( \psi_\tau(\tilde{r}(\xho)) \right)\hat{\sigma}_M(r(\xho))\am^\top\right]\bigg),
\label{eq:infFunction1}
\end{align} 
where $\psi_\tau(u)$ is defined as
\begin{eqnarray}
\psi_\tau (\xheps) &=& \nonumber \\  
 & & \!\!\!\!\!\!\!\!\!\!\!\!\!\!\!\!\!\!\!\!\!\!  w(\xheps)\psi_1(\tilde{r}(\xheps))  +  \psi_2 (\tilde{r}(\xheps)), \nonumber \\  
\end{eqnarray}
with weights $w(\xheps)$ given by
\begin{eqnarray}
w(\xheps) :=  \nonumber \\  
 & & \!\!\!\!\!\!\!\!\!\!\!\!\!\!\!\!\!\!\!\!\!\!\!\!\!\!\!\!\!\!\!\!\!\!\!\!\!\!   \frac{\mathbb{E}_{H_\epsilon} [ 2\rho_2(\tilde{r}(\xheps))  -  \psi_2(\tilde{r}(\xheps))\tilde{r}(\xheps)]}{\mathbb{E}_{H_\epsilon}
[\psi_1(\tilde{r}(\xheps)) \tilde{r}(\xheps)]},  \nonumber \\  
\end{eqnarray}
and the partial derivative $\frac{\partial (\psi_\tau(\tilde{r}(\xho))\hat{\sigma}_M(r(\xho)))}{\partial \xh^{\tau_R}}$ is given by Equation (\ref{eq:mainPartial}) in Appendix \ref{app:IFder}.
\label{th:theorem1}
\end{theorem}

The detailed proof of the theorem is in Appendix \ref{app:IFder}. The main idea of the proof is to notice that the estimate $\xh^{\tau_R}(H)$ should minimize (\ref{eq:asTau}), so it should satisfy 
\begin{align}
-\mathbb{E}_{H_\epsilon} [ \psi_\tau (\xheps)\hat{\sigma}_M(r(\xheps))\am^\top] 
+\lambda J'(\xh^{\tau_R}) = 0.
\label{eq:FE}
\end{align} 
Deriving this expression w.r.t. $\epsilon$, and particularizing for $\epsilon=0$, we arrive to Equation (\ref{eq:infFunction1}).

But the IF given in Theorem \ref{th:theorem1} is not valid for a non-differentiable regularization function like $J(x) = |x|$. We should find the IF for this case using a different technique: we approximate $|x|$ with a twice differentiable function $J_K(x)$ such that
\begin{align}
\lim_{K\to\infty} J_K(x) = |x| .
\end{align}
In particular, we choose
\begin{align}
J_K(x) = x\tanh (Kx).
\end{align}

The IF of a $\tau$ estimator regularized with $J_K$, denoted by $\xh^{\tau_K}$, is given in Theorem \ref{th:theorem1}. The limit of this IF when $K\to\infty$ is the IF for the $\tau$ estimator regularized with $|x|$, denoted by $\xh^{\tau_{\ell_1}}$
\begin{align}
\lim_{K\to\infty} \text{IF}((\am_0,y_0), \xh^{\tau_K}, H_0) = \text{IF}((\am_0,y_0), \xh^{\tau_{\ell_1}}, H_0).
\end{align}
This idea leads us to the next theorem.
\begin{theorem}
Let $y = \am\x_0 + e$ be as given in (\ref{eq:basicModel}). Let also $\rho_1,\rho_2$ be twice differentiable functions. Assume the M-scale $\hat{\sigma}_M(u)$ as given in (\ref{eq:robscaleAs1}). Define $\tilde{r}(\x):=r(\x)/\hat{\sigma}_M(r(\x)) $ and $\psi_i(u) = \rho'_i(u)$. Without loss of generality, assume the regularized $\tau$ estimate to be t-sparse, $\xh^{\tau_R}(H) = (\xh^{\tau_R}_1(H),\dots,\xh^{\tau_R}_t(H),0,\dots,0)$. If $J(x)=|x|$, then the IF for the regularized $\tau$ estimator $\xh^{\tau_R}$ is given by 

\begin{align}
&\text{IF}((\am_0,y_0), \xh^{\tau_R}, H_0)= \nonumber \\
&\begin{pmatrix}
&\left(-\mathbb{E}_{H_0} [\am_{1:t}^\top(\frac{\partial \psi_\tau(\tilde{r}(\xho))\hat{\sigma}_M(r(\xho))}{\partial \xh^{\tau_R}})_{1:t}]\right)^{-1} \\ 
& \Big(\psi_\tau\left(\frac{y_0 - \am_0\xho}{\hat{\sigma}_M(r(\xho))}\right)_{\! 1:t}\! \hat{\sigma}_M(r(\xho))(\am^\top_0)_{1:t}\\
&\mathbb{-E}_{H_0}\left[\left( \psi_\tau(\tilde{r}(\xho)) \right)_{1:t}\hat{\sigma}_M(r(\xho))\am_{1:t}^\top\right]\Big)\\ 
&\mathbf{0}_{n - t}
 \end{pmatrix}, 
 \label{eq:ifL1}
\end{align}
\label{th:theorem2}

where  $\frac{\partial (\psi_\tau(\tilde{r}(\xho))\hat{\sigma}_M(r(\xho)))}{\partial \xh^{\tau_R}}$, $\psi_\tau(u)$ and $w(\xheps)$ are defined as in Theorem\ref{th:theorem1} and $\x_{1:t}$ refers to the column vector containing the first $t$ elements of $\x$.
\end{theorem}

For more details about the proof, see Appendix \ref{app:IF}. Surprisingly, in this case the IF does not depend on the regularization parameter $\lambda$.

The IFs given by Theorems 1 and 2 are not bounded in $a_0$, but they are bounded in $y_0$ if $\psi_\tau$ is bounded (i.e., if $\psi_1$ and $\psi_2$ are bounded). This also holds for the original non-regularized $\tau$ estimator. We assert then that the regularization does not change the robustness properties of the $\tau$ estimator, as measured by the IF. 

\subsubsection{Examples of influence functions}
 
In the high-dimensional case, the IF cannot be plotted. So, to visualize the results that we obtained in Theorems \ref{th:theorem1} and \ref{th:theorem2}, we create 1-dimensional  examples.  

We model $a$ and the additive errors $e$ as Gaussian random variables with zero mean and variance equal to one. The ground truth $x_0$ is fixed to $1.5$. We use $\rho_1$ and $\rho_2$ from the optimal family \cite{salibian08}, with clipping parameters $c_1=1.21$ and $c_2=3.27$ (see Figure \ref{fig:rhosopt}). The regularization parameter $\lambda$ is set to 0.1. 
\begin{figure*}[ht!]
\centering
  \includegraphics[width=1\linewidth]{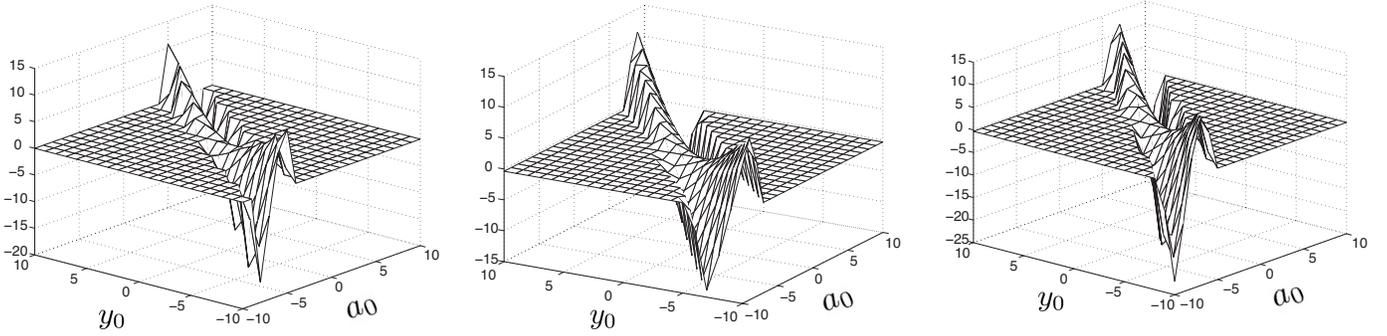}
  \caption{IF of the $\tau$ estimators for a 1-dimensional example. The left plot shows the IF for the non-regularized $\tau$ estimator. The middle plot depicts is the IF for the $\ell_2$-regularized estimator with $\lambda=0.1$. The right plot provides the IF for the $\ell_1$-regularized $\tau$ estimator with $\lambda=0.1$.}
  \label{fig:IFtau}
\end{figure*}

In this setup, we compute the IF for the non-regularized, the $\ell_2$-regularized, and the $\ell_1$-regularized $\tau$ estimators.  Figure \ref{fig:IFtau} shows the resulting IFs. We can observe that in all cases the IF is bounded both in $a_0$ and $y_0$, so we can consider that in this example the estimators are robust against small fractions of contaminations in $a$ and in $e$. Also, we notice that the amplitude of the IF is similar for the three different cases. This tells us that the introduction of the regularization does not affect the sensitivity of the estimate to small contaminations of the data.

\subsubsection {Sensitivity curves}

The SC is the finite sample version of the IF. It was initially proposed by J.W. Tukey \cite{huber02}. Following \cite{huber02}, we define the SC as
\begin{equation}
\text{SC}(\am_0,y_0,\xh_m) = \frac{\xh_m(\A\cup\{\am_0\},\y \cup \{y_0\}) - \xh_m(\A,\y))}{1/(m+1)}.
\end{equation}

We compute the SCs corresponding to the IF examples. We use $m=1000$ samples taken from the populations described above. The outliers $a_0$ and $y_0$ range from -10 to 10. The estimates $\xh_m^\tau(\A\cup\{a_0\},\y \cup \{y_0\}$ and $\xh_m^\tau(\A,\y))$ are computed using the fast algorithms described in Algorithm \ref{alg:fast}.

\begin{figure*}[ht!]
\centering
  \includegraphics[width=1\linewidth]{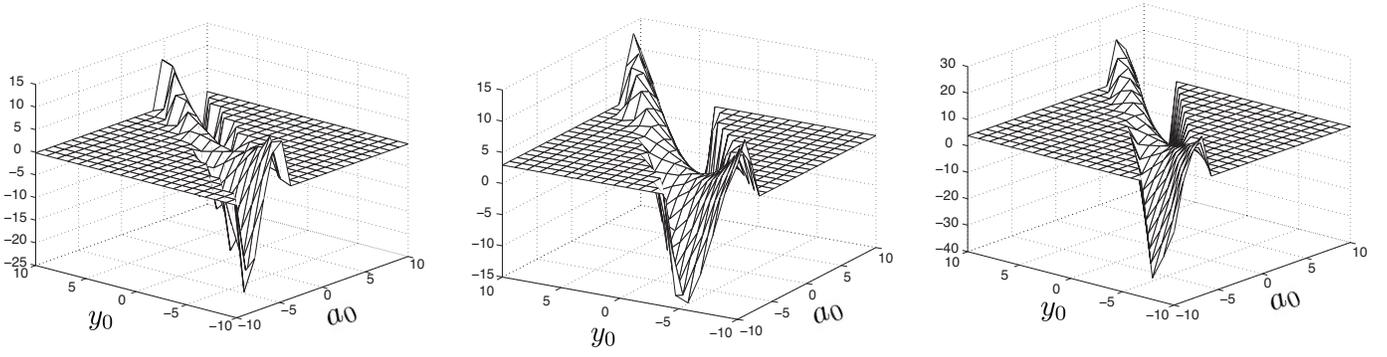}
  \caption{SC of the $\tau$ estimators. On the left is the SC for the non-regularized $\tau$ estimator. On the center is the SC for the $\ell_2$-regularized estimator with $\lambda=0.1$. On the right is the SC for the $\ell_1$-regularized $\tau$ estimator with $\lambda=0.1$.}
  \label{fig:SC}
  \vspace{-0.8em}
\end{figure*}

The resulting SCs for the non-regularized, $\ell_2$ and $\ell_1$ regularized estimators are shown in Figure \ref{fig:SC}. In all the cases the SC matches closely its corresponding IF. This also gives us an indication of the performance of the proposed algorithms: although there is no guarantee to find the global minimum, the computed estimates, at least in this particular example, are close to their theoretical values.

\subsection{Many outliers}
So far we have studied the robustness of the regularized $\tau$ estimators when the proportion of outliers in the data is very small. Also, we have explored the behaviour of the estimators in low dimensional problems. The goal now is to investigate how the estimators empirically behave when there are more outliers in the data (up to 40 $\%$) and when the dimension of the problem is higher. This will give us an estimate of the brekdown point of the different estimators.

One reasonable requirement for our estimators is to have simultaneously a small bias and a small variance. This can be measured by the MSE
\begin{align}
\text{MSE}(\xh,H)& = \mathbb{E}_{H}[|\!|\xh - \x_0|\!|^2 ]  \\
& = \text{Tr }\{\text{Var}(\xh,H)\} + |\!|\text{Bias}(\xh,H)|\!|^2 ,  
\end{align}
where the bias of the estimator $\xh$ at distribution $H$ is
\begin{equation}
\text{Bias}(\xh,H) = \mathbb{E}_{H}[\xh] - \x_0,
\end{equation}
and its covariance matrix is
\begin{equation}
\text{Var}(\xh,H) = \mathbb{E}_{H}[(\xh - \mathbb{E}_{H}[\xh])(\xh - \mathbb{E}_{H}[\xh])^\top].
\end{equation}
The sample version of the MSE is defined as
\begin{align}
\text{MSE}(\xh,H) = \frac{1}{m}\sum_{i=1}^m (\xh - \x_0)^2.
\end{align}

We use these metrics to summarize and compare the performance of different estimators. 

We focus on the  non-regularized $\tau$ estimator and the $\ell_2$ and $\ell_1$ regularized $\tau$ estimators.
We compare them with other estimators that use the same regularizations, but different loss functions.
In particular, we compare them with estimators that use LS and M loss functons. In the case of the M loss function, a key difficulty is to estimate the scale of the errors $\e$ (see Equation (\ref{eq:mestimator})). Depending on the quality of this scale estimate, the performance of the M estimators changes significantly. This is why we generate two different M estimates: The first one uses a preliminary estimate of the scale. This estimate is computed using the median absolute deviation \cite{Rousseeuw87} applied to the residuals generated by the LS estimate. The second one uses the ground truth value of $\sigma$ in Eq.~(\ref{eq:mestimator}) to scale the residuals in the M loss function. In fact, this estimator also corresponds to an MM estimator with perfect S-step. It is thus referred to as "MM opt scale" and provides an upper bound on the possible performance of an MM estimator. Notice that the $\tau$ estimator does not require any preliminary estimation of the scale. The $\rho$ function in the M-estimator is Huber's function \cite{huber09}, while in the $\tau$-estimator we choose $\rho_1$ and $\rho_2$ to be optimal weight functions \cite{salibian08}, shown in Figure \ref{fig:rhosopt}.

To perform the study, we run numerical simulations. The setup of the simulations is slightly different, depending on the type of regularization that we use. Our main goal here is to observe the deviations produced by outliers. We avoid other effects that could mask the outliers. That is why we satisfy the different assumptions in each regularization.

\subsubsection{Common settings for all simulations}
All the experiments share the following settings:
We generate a matrix $\A\in\R^{60\times 20}$ with random i.i.d. Gaussian entries. The measurements $\y$ are generated using additive outliers and additive Gaussian noise 
 
\begin{equation}
\y = \A\x + \e_G + \e_o,
\end{equation}
where $\e_{G}\in\R^{60\times 1}$ has random i.i.d. Gaussian entries. On the other hand, $\e_o\in\R^{60\times 1}$ is a sparse vector. The few non-zero randomly selected entries have a random Gaussian value with zero mean and variance equal to 10 times the variance of the noiseless data. These entries represent the outliers. For each experiment, we perform 100 realizations. In the case of the regularized estimators, we manually select the optimum regularization parameter $\lambda$ with respect to the MSE.
We carry out experiments with 0\%, 10\%, 20\%, 30\%, 40\% of outliers in the data.
\begin{figure*}[ht!]
\centering
  \includegraphics[width=1\linewidth]{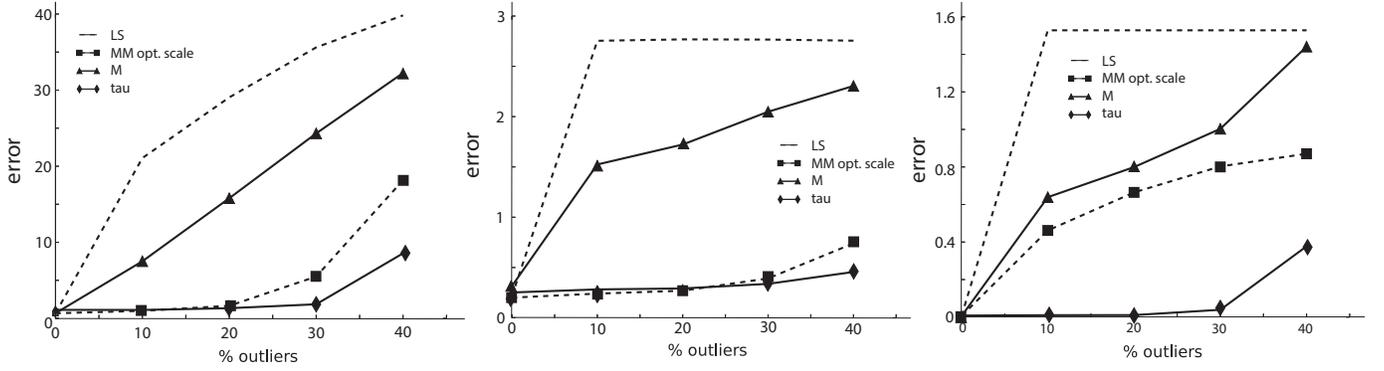}
  \caption{Results of the experiments to explore and compare the behaviour of the $\tau$ estimators in the presence of different proportions of additive outliers. The plot on the left corresponds to non-regularized estimators, the plot in the center corresponds to $\ell_2$-regularized estimators and the plot in the right corresponds to $\ell_1$-regularized estimators.}
  \label{fig:exp}
  \vspace{-0.8em}
\end{figure*}

\subsubsection{Specific settings in each simulation}
The experiments where the non-regularized estimators are compared use a dense source $\x$ and a model matrix $\A$ with a low condition number of the order of 10. The experiments where the $\ell_2$ regularized estimators are compared use a dense source and model matrix with a  condition number of 1000. The experiments where the $\ell_1$ regularized estimators are compared use a sparse source (20\% of non zero entries) and a model matrix with a condition number of 1000.

\subsubsection{Results}
The results of the experiments are given in Figure \ref{fig:exp}. They are grouped by regularization type. In the first experiment, as the non-regularized LS, M, MM opt scale, and $\tau$ estimators are unbiased, the MSE is equivalent to the variance of the estimators.
With clean data ($0\%$ outliers) the variance of the $\tau$ estimator is slightly larger than the variance of the other estimators. 
However, the $\tau$ estimator performs better when there are more outliers in the data, it even outperforms the MM opt scale estimator when $30-40\%$ of the data is contaminated.

In the second experiment ($\ell_2$-regularized estimators), we find a similar behaviour as in the first experiment: the $\tau$ estimator is slightly worse than the M, MM opt scale, and LS estimators when the data is not contaminated, but it surpases the performance of all the others when $30-40\%$ of the data are outliers.
Since we select the optimal regularization $\lambda$, there is a bound on the error of all the regularized estimators. In other words, if we increase $\lambda$ sufficiently, we will force the algorithm to return the $\xh = 0$ solution. That is the worse case solution, corresponding to a breakdown, and we can see it clearly in the LS estimator.

In the third experiment ($\ell_1$-regularized estimators), for all cases with $>0\%$ outliers, the $\tau$ estimator significantly outperforms all other estimators. 

\subsection{Clean data}

In this section we study the behaviour of the regularized $\tau$ estimator when there are no outliers in the data, i.e. when the errors are Gaussian: $H = H_0$. In particular, we study how the introduction of the regularization affects the bias and the asymptotic variance of the estimator.

The asymptotic variance (ASV) is defined as
\begin{equation}
\text{ASV}(\xh,H) = m \lim_{m\to\infty}\text{Var}(\xh_m),
\end{equation}
We explore it in a one-dimensional example with numerical simulations. In this case, the source $x_0$ is set to 1.5, the model $a$ is a standard Gaussian random variable and the measurements $y$ are generated adding standard Gaussian errors $e$. We use $m = 5000$ measurements.

\begin{figure}[ht!]
\centering
  \includegraphics[width=0.9\linewidth]{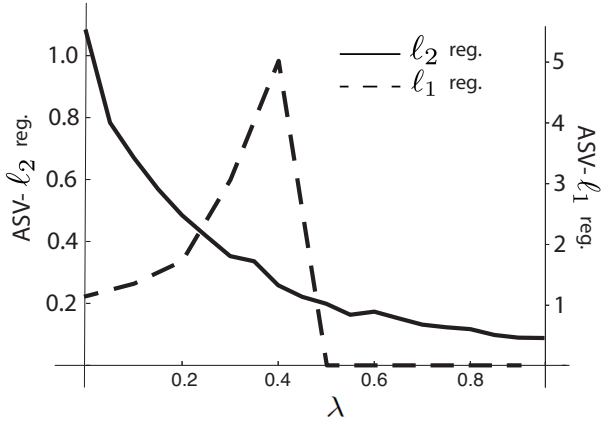}
  \caption{The figure represents the ASV for the $\ell_2$ and $\ell_1$ regularized estimator in a one-dimensional problem.}
  \label{fig:ASV}
\end{figure}

The upper part of Figure \ref{fig:ASV} shows the ASV of the $\ell_2$- regularized $\tau$ estimator for different values of the parameter $\lambda$. We can observe that, as $\lambda$ grows, the ASV of this estimator decreases. However, in the $\ell_1$ regularized case, the opposite happens. It is shown in the lower part of Figure \ref{fig:ASV}: as the value of $\lambda$ increases, the ASV of the estimator increases as well.

\begin{figure}[ht!]
\centering
  \includegraphics[width=.9\linewidth]{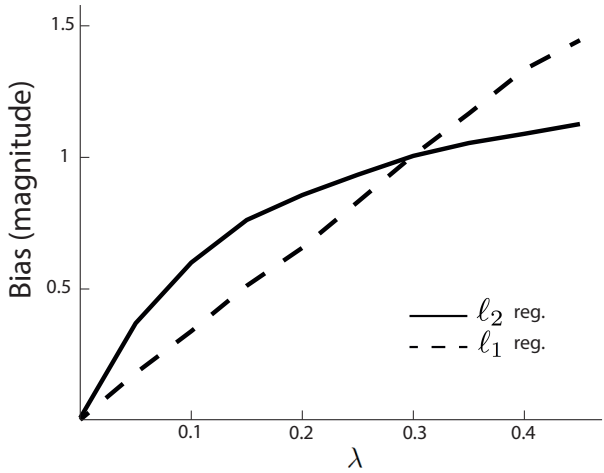}
  \caption{Bias of the regularized $\tau$ estimator in a one-dimensional problem. The solid line corresponds to an $\ell_2$ regularized estimator. The dashed line corresponds to an $\ell_1$ regularized estimator.}
  \label{fig:bias}
  \vspace{-1.2em}
\end{figure}

The non-regularized $\tau$ estimator has zero bias \cite{yohai88}, but the introduction of the regularization biases the estimator. To study this bias in the $\ell_2$ and $\ell_1$ regularized cases, we again use the one-dimensional example described above. We compute  the expectation $\mathbb{E}_{H}[\xh_\infty(H)]$ using Monte Carlo integration. Results are shown in Figure \ref{fig:bias}. With both regularizations the magnitude of the bias increases with $\lambda$. 

%

\section{Conclusions}
\label{sec:conclusion}
We proposed a new robust regularized estimator that is suitable for linear inverse problems that are ill-posed and in the presence outliers in the data.
We also proposed algorithms to compute these estimates. Furthermore, we provided an analysis of the corresponding estimators. We studied their behaviour in three different situations: when there is an infinitesimal contamination in the data, when the data is not contaminated, and when there are many outliers in the data. For that, we derived their influence function, and computed their sensitivity curve, bias, variance, and MSE. We showed for a 1-dimensional example that the error of the regularized $\tau$ estimate is bounded for additive outliers: a single outlier, regardless of its magnitude, cannot lead to an infinite deviation in the estimate. The ASV of the $\ell_2$ regularized estimator decreased with the regularization parameter $\lambda$. In the $\ell_1$ case, it increased with $\lambda$. The magnitude of the bias increased with $\lambda$ when we introduced the regularizations. In higher dimensional simulation examples, the regularized $\tau$ estimators had a smaller MSE in the presence of outliers compared to the regularized LS, M and MM estimators. This was especially true in the $\ell_1$ regularized case. Future work will consider applying the proposed estimator to the challenging problem of estimating the temporal releases of a pollutant to the atmosphere using temporal measurements of the concentration of the pollutant in the air taken at different locations.


%

\appendices
\section{Finding local minima }\label{ap:localminima}
Every local minimum satisfies
\begin{equation}
\frac{\partial (\hat{\sigma}_\tau^2(\rr(\x)) + \lambda \sum_{i=1}^n J(x_i))}{\partial \x} = 0.
\end{equation}
Let us develop this expression. For clarity, we first define
\begin{equation}
\tilde{\rr}(\x) := \frac{\rr(\x)}{\hat{\sigma}_M(\rr(\x))}, \quad \psi_j (\x):= \frac{\partial \rho_j(\x)}{\partial \x}.
\end{equation}
Using the definitions of $\hat{\sigma}_\tau(\rr(\x))$ in Equation (\ref{eq:tausigma}) and $\hat{\sigma}_M (\rr(\x))$ in Equation (\ref{eq:robscale}), we get
\begingroup\makeatletter\def\f@size{9}\check@mathfonts
\begin{align}
&\frac{\partial (\hat{\sigma}_\tau^2(\rr(\x)) + \lambda \sum_{i=1}^n J(x_i))}{\partial \x}  =\\ &2\hat{\sigma}_M (\rr(\x))\frac{\partial{\hat{\sigma}_M(\rr(\x))}}{\partial{\x}} \frac{1}{m}\sum_{i=1}^m \rho_2(\tilde{r}_i(\x)) + \nonumber \\
&+  \hat{\sigma}_M^2 (\rr(\x))\frac{1}{m} \sum_{i=1}^m \psi_2(\tilde{r}_i(\x))\left[ \frac{-\am^\top_i\hat{\sigma}_M(\rr(\x)) - r_i(\x) \frac{\partial{\hat{\sigma}_M (\rr(\x))}}{\partial{\x}}}{\hat{\sigma}_M^2 (\rr(\x))}\right] \nonumber\\
 &+  \lambda J'(\x) = 0,
\label{eq:tauDer}
\end{align} 
\endgroup
where $\am_i$ is the $i$th row of the matrix $\A$ and $J'(\x)$ is an abbreviation for $(\frac {\partial J(\x)}{\partial x_1}, \dots, \frac {\partial J(\x)}{\partial x_n})^\top$.

To find $\frac{\partial{\hat{\sigma}_M (\rr(\x))}}{\partial{\x}}$, we take the derivative of (\ref{eq:robscale}) w.r.t. $\x$:
\begin{align}
\frac{\partial{\hat{\sigma}_M(\rr(\x))}}{\partial{\x}} = -\hat{\sigma}_M(\rr(\x)) \frac{\sum_{i=1}^m \psi_1 (\tilde{r}_i(\x))\am_i^\top}{\sum_{i=1}^m \psi_1 (\tilde{r}_i(\x))r_i(\x)}
\label{eq:scaleDer1}
\end{align}
Replacing (\ref{eq:scaleDer1}) in (\ref{eq:tauDer}), we obtain
\begin{align}
-\frac{1}{m} \sum_{i=1}^m\left( w_m(\x)\psi_1(\tilde{r}_i(\x))  +  \psi_2 (\tilde{r}_i(\x))\right)&\hat{\sigma}_M(\rr(\x))\am^\top_i +\nonumber \\
&+\lambda J'(\x) = 0,
\label{eq:fundEq}
\end{align} 
where 
\begin{align}
w_m(\x) :=  \frac{\sum_{i=1}^m 2\rho_2(\tilde{r}_i(\x))  - \sum_{i=1}^m \psi_2(\tilde{r}_i(\x))\tilde{r}_i(\x)}{\sum_{i=1}^m
\psi_1(\tilde{r}_i(\x)) \tilde{r}_i(\x)}.  
\end{align}

We can see that (\ref{eq:fundEq}) is also the derivative of a penalized reweighted least squares function 
\begin{align}
f(\x) = \frac{1}{m}\sum_{i=1}^m z(\tilde{r}_i(\x)) (y_i - \am_i\x)^2 + \lambda J(\x)
\label{rwls}
\end{align}
with weights
\begin{align}
z(u) = \begin{cases} \frac{\psi_{\tau}(u)}{2u} & \quad \text{if } u\neq0 \\ 
0 & \quad \text{otherwise, } \\ \end{cases} 
\label{eq:weights}
\end{align}
where
\begin{align}
\psi_{\tau}(u) = w_m(\x)\psi_1(u) +  \psi_2 (u).
\end{align}
So $f(\x)$ has the same minium as (\ref{eq:regtau}).

\subsection{$\ell_2$ regularization}
When $J(x) = x^2$, $J(x)$ is differentiable. Then, we can apply the theory from above and, hence minimizing
\begin{equation}
\hat{\sigma}^2_\tau(\rr(\x)) + \lambda\sum_{i=1}^n x_i^2
\label{eq:ridgetau}
 \end{equation}
is equivalent to minimizing
\begin{equation} 
\norm{\ZZ^{1/2}(\x)(\A\x - \y)}^2_2 + \lambda \sum_{i=1}^n x_i^2.
\label{eq:WLS2}
\end{equation}

\subsection{$\ell_1$ regularization}
In the case of $J(x) = |x|$ , $J(x)$ is not differentiable. To overcome this problem, we approximate $|x|$ with a differentiable function. We chose
\begin{align}
J_K(x) &= x\tanh (Kx),
\end{align}
where
\begin{align}
\lim_{K\to\infty} J_K(x) &= |x|.
\end{align}
Then, we perform the same derivation as in the last section, but using $J_K (x)$. In this case, Equation (\ref{rwls}) becomes
\begin{equation}
f_K(\x) = \sum_{i=1}^m w_i(\x) (y_i - \am_i\x)^2 + \lambda\sum_{i = 1}^n J_K(x_i).
\label{rwlsk}
\end{equation}
Now, taking limits
\begin{equation}
f(\x) = \lim_{K\to\infty} f_K(\x) = \sum_{i=1}^m w_i(\x) (y_i - \am_i\x)^2 + \lambda\sum_{i = 1}^n |x_i|
\label{rwlsL1}
\end{equation}
So minimizing (\ref{eq:regtau}) with $J(x) = |x|$  is equivalent to minimizing
\begin{equation}
f_1(\x) = \sum_{i=1}^m w_i(\x) (y_i - \am_i\x)^2 + \lambda\sum_{i = 1}^n |x_i|.
\label{rwls1}
\end{equation}

\section{Influence Function of the $\tau$ estimator with twice differentiable $J(x)$}\label{app:IFder}
\begin{proof}
From Equation (\ref{eq:fundEq}), we know that, at the contaminated distribution $H_\epsilon$, the estimate has to satisfy
\begin{align}
-\mathbb{E}_{H_\epsilon} [ \psi_\tau (\xheps)\hat{\sigma}_M(r(\xheps))\am^\top] 
\!+\! \lambda J'(\xheps)\! = \! 0
\label{eq:FE}
\end{align} 
where
\begin{eqnarray}
\psi_\tau (\xheps) &=& w(\xheps)\psi_1(\tilde{r}(\xheps)) \nonumber \\ 
				      & & +  \psi_2 (\tilde{r}(\xheps)), 
\end{eqnarray}
with
\begin{eqnarray}
w(\xheps) &:=& \nonumber \\
& & \!\!\!\!\!\!\!\!\!\!\!\!\!\!\!\!\!\!\!\!\!\!\!\!\!\!\!\!\!\!\!\!  \frac{\mathbb{E}_{H_\epsilon} [ 2\rho_2(\tilde{r}(\xheps))  -  \psi_2(\tilde{r}(\xheps))\tilde{r}(\xheps)]}{\mathbb{E}_{H_\epsilon}
[\psi_1(\tilde{r}(\xheps)) \tilde{r}(\xheps)]}.  \nonumber \\
\end{eqnarray}

\begin{equation}
\mathbb{E}_{H_\epsilon} \left[ \rho_1\left(\frac{r(\xheps)}{\hat{\sigma}_M(r(\xheps))}\right)\right] = b, 
\label{eq:robscaleAs}
\end{equation}
and $b = E_{H_0}[\rho_1(u)]$. Using the definition of the contaminated distribution $H_\epsilon = (1-\epsilon) H_0 +\epsilon\delta_{\am_0,y_0} $, we can rewrite (\ref{eq:FE}) as
\begin{align}
-(1 - \epsilon)\mathbb{E}_{H_0} [\left( \psi_\tau(\tilde{r}(\xheps)) \right)&\hat{\sigma}_M(r(\xheps))\am^\top]  \nonumber\\ 
-\epsilon \left( \psi_\tau\left(\frac{y_0 - \am_0\xheps}{\hat{\sigma}_M(r(\xh^{\tau_R}))}\right)\right)&\hat{\sigma}_M(r(\xheps))\am_0^\top  \nonumber\\ 
&+\lambda J'(\xheps) = 0.
\label{eq:derivative}
\end{align} 

Taking derivatives w.r.t. $\epsilon$
\begin{align}
\quad \mathbb{E}_{H_0}& [\left( \psi_\tau(\tilde{r}(\xheps)) \right)\hat{\sigma}_M(r(\xheps))\am^\top]  \nonumber\\ 
&\!\!\!\!\!\!\!\!\!\!-(1 - \epsilon)\mathbb{E}_{H_0} \left[\am^\top\frac{\partial \left(\psi_\tau(\tilde{r}(\xheps)) \hat{\sigma}_M(r(\xheps))\right)}{\partial \epsilon} \right] \nonumber\\ 
&\!\!\!\!\!\!\!\!\!\! - \epsilon\bigg( \frac{\partial(\psi_\tau\left(\frac{y_0 - \am_0\xheps}{\hat{\sigma}_M(r(\xheps))}\right)\hat{\sigma}_M(r(\xheps))\am_0^\top )}{ \partial \epsilon}\bigg)  \nonumber\\
&- \psi_\tau\left(\frac{y_0 - \am_0\xheps}{\hat{\sigma}_M(r(\xheps))}\right)\hat{\sigma}_M(r(\xheps))\am_0^\top \nonumber \nonumber\\ 
&+\lambda J''(\xheps)\frac{\partial\xheps}{\partial\epsilon} = 0,
\label{eq:derivative2}
\end{align} 
where $J''(\xh^{\tau_R})$ represents the Jacobian matrix.
Using the chain rule for derivation, we get
\begin{align}
&\frac{\partial (\psi_\tau(\tilde{r}(\xheps)) \hat{\sigma}_M(r(\xheps)))}{\partial\epsilon}= \nonumber\\ 
&\frac{\partial (\psi_\tau(\tilde{r}(\xheps)) \hat{\sigma}_M(r(\xheps)))}{\partial\xheps}
\frac{\partial \xheps}{\partial\epsilon}
\end{align}
where
\begin{align}
&\frac{\partial (\psi_\tau(\tilde{r}(\xheps)) \hat{\sigma}_M(r(\xheps)))}{\partial\xheps}=\nonumber\\ 
&\frac{\partial\psi_\tau(\tilde{r}(\xheps))}{\partial \xheps} \hat{\sigma}_M(r(\xheps))\nonumber\\ 
&+\psi_\tau(\tilde{r}(\xheps))\frac{\partial\hat{\sigma}_M(r(\xheps))}{\partial\xheps}
\label{eq:mainPartial}
\end{align}
and
\begin{align}
\frac{\partial \psi_\tau(\tilde{r}(\xheps))}{\partial \xheps}=\frac{\partial\psi_\tau(\tilde{r}(\xheps))}{\partial \tilde{r}(\xheps)}\frac{\partial \tilde{r}(\xheps)}{\partial \xheps}.
\label{eq:partialscore}
\end{align}
We already know from (\ref{eq:scaleDer1}) that
\begin{align}
\frac{\partial{\hat{\sigma}_M(r(\xheps))}}{\partial{\xheps}} = \nonumber \\
&\!\!\!\!\!\!\!\!\!\!\!\!\!\!\!\!\!\!\!\!\!\!\!\!\!\!\!\!\!\!\!\!\!\!\!\!\!\!-\hat{\sigma}_M(r(\xheps)) \frac{\mathbb{E}_{H_\epsilon} [\psi_1 (\tilde{r}(\xheps))\am^\top]}{\mathbb{E}_{H_\epsilon}[\psi_1 (\tilde{r}(\xheps))r(\xheps)]}.
\label{eq:scaleDer}
\end{align}
We also need
\begin{align}
&\frac{\partial\psi_\tau(\tilde{r}(\xheps))}{\partial \tilde{r}(\xheps)} = \nonumber\\ 
&\frac{\partial w(\tilde{r}(\xheps))}{\partial \tilde{r}(\xheps)}\psi_1(\tilde{r}(\xheps)) \nonumber\\ 
&+ w(\tilde{r}(\xheps))\psi'_1(\tilde{r}(\xheps)) +\psi'_2(\tilde{r}(\xheps))
\end{align}
that uses
\begin{align}
&\frac{\partial w(\tilde{r}(\xheps))}{\partial \tilde{r}(\xheps)} = \nonumber\\ 
& \frac{\mathbb{E}_{H_\epsilon}[2\psi_2(\tilde{r}(\xheps)-2\rho_2(\tilde{r}(\xheps)]}{\mathbb{E}_{H_\epsilon}[\psi_1(\tilde{r}(\xheps)) \tilde{r}(\xheps)]^2}\nonumber \\
&- \frac{\mathbb{E}_{H_\epsilon}[\psi'_2(\tilde{r}(\xheps))\tilde{r}(\xheps)]}{\mathbb{E}_{H_\epsilon}[\psi_1(\tilde{r}(\xheps)) \tilde{r}(\xheps)]^2} \nonumber\\ 
&- \frac{\mathbb{E}_{H_\epsilon}[\psi_2(\tilde{r}(\xheps)))\psi_1(\tilde{r}(\xheps)) \tilde{r}(\xheps)]}{\mathbb{E}_{H_\epsilon}[\psi_1(\tilde{r}(\xheps)) \tilde{r}(\xheps)]^2} \nonumber\\ 
& -\frac{\mathbb{E}_{H_\epsilon}[-  \psi_2(\tilde{r}(\xheps))\tilde{r}(\xheps))(\psi_1'(\tilde{r}(\xheps)) \tilde{r}(\xheps)+ \psi_1(\tilde{r}(\xheps)) )]
}{\mathbb{E}_{H_\epsilon}[\psi_1(\tilde{r}(\xheps)) \tilde{r}(\xheps)]^2} 
\end{align}
In Equation (\ref{eq:partialscore}), we also need
\begin{align}
\frac{\partial \tilde{r}(\xheps)}{\partial \xheps}= \frac{-\am^\top\hat{\sigma}_M(r(\xheps)) - r(\xheps)\frac{\partial \hat{\sigma}_M(r(\xheps))}{\partial \xheps}}{\hat{\sigma}^2_M(r(\xheps))}
\end{align}
We have now all the elements of Equation (\ref{eq:derivative2}). The next step, in order to find the IF, is to particularize Equation (\ref{eq:derivative2}) for $\epsilon = 0$. Since $\frac{\partial\xheps}{\partial\epsilon}|_{\epsilon=0}=\text{IF}((\am_0,y_0), \xh^{\tau_R}, H_0)$, we can write
\begin{align}
\mathbb{E}_{H_0}& [ \psi_\tau(\tilde{r}(\xho))\hat{\sigma}_M(r(\xho))\am^\top]  \nonumber\\ 
&- \mathbb{E}_{H_0} \left[\am^{\top}\frac{\partial \psi_\tau(\tilde{r}(\xho))\hat{\sigma}_M(r(\xho))}{\partial r(\xho)}
\frac{\partial r(\xho)}{\partial \xho}\right] \nonumber\\ 
&\times \text{IF}((\am_0,y_0), \xh^{\tau_R}, H_0) \nonumber\\ 
&- \left( \psi_\tau\left(\frac{y_0 - \am_0\xho}{\hat{\sigma}_M(r(\xho))}\right)\right)\hat{\sigma}_M(r(\xho))\am_0^\top\nonumber \\
&+\lambda J''(\xho)\text{IF}((\am_0,y_0), \xh^{\tau_R}, H_0) = 0.
\label{eq:derivative}
\end{align} 
Solving the last equation for $\text{IF}((\am_0,y_0), \xh^{\tau_R}, H_0)$, we get
\begin{align}
&\text{IF}((\am_0,y_0), \xh^{\tau_R}, H_0) = \\ \nonumber
&\left(-\mathbb{E}_{H_0} [\am^\top\frac{\partial \psi_\tau(\tilde{r}(\xho))\hat{\sigma}_M(r(\xho))}{\partial \xho}] +\lambda J''(\xho)\right)^{-1} \nonumber\\ 
&\bigg( \psi_\tau\left(\frac{y_0 - \am_0\xho}{\hat{\sigma}_M(r(\xho))}\right)\hat{\sigma}_M(r(\xho))\am_0^\top\nonumber\\ 
&\mathbb{-E}_{H_0}\left[\left( \psi_\tau(\tilde{r}(\xho)) \right)\hat{\sigma}_M(r(\xho))\am^\top\right]\bigg)
\label{eq:infFunction}
\end{align} 
\end{proof}
\section{Influence Function for the $\tau$ estimator with $\ell_1$ regularization}\label{app:IF}
\begin{proof}
We approximate $J(x) = |x|$ with a twice differentiable function
\begin{align}
J_K(x) &= x\tanh (Kx) \nonumber \\
\lim_{K\to\infty} J_K(x) &= |x| .
\end{align}
The second derivative of $J_K(x)$ is
\begin{align}
J_K''(x) = &2K(1-\tanh^2(Kx)) - \nonumber \\
& 2K^2x\tanh(Kx)(1 - \tanh^2(Kx)).
\end{align}
In particular, 
\begin{equation}
J_K''(0) = 2K.
\end{equation}
For $J(x)$ twice differentiable, we know that the IF of a regularized $\tau$ estimator is (\ref{eq:infFunction}). Without loss of generality, we can assume that the estimate $\xho$ is $t$-sparse. Then,
\begin{align}
\text{IF}&((\am_0,y_0), \xh^{\tau_R}, H_0) = \nonumber \\
&(-\mathbb{E}_{H_0} \left[\am^\top\frac{\partial ( \psi_\tau(\tilde{r}(\xho))\hat{\sigma}_M(r(\xho)))}{\partial \xho}\right] \nonumber  \\
 & +2\lambda \text{diag} ((J''(\xh^{\tau_R(H_0)}))_1,\dots(J''(\xh^{\tau_R(H_0)}))_t,2K,\dots,2K))^{-1} \nonumber \\ \nonumber
&+ \left( \psi_\tau\left(\frac{y_0 - \am_0^\top\xho}{\hat{\sigma}_M(r(\xho))}\right)\right)\hat{\sigma}_M(r(\xho))\am_0^\top\\ \nonumber
&\mathbb{-E}_{H_0}\left[\left( \psi_\tau(\tilde{r}(\xho)) \right)\hat{\sigma}_M(r(\xho))\am^\top\right].
 \end{align}

Now, we focus on the inversion of the matrix. For that, let us work with block matrices. If we call

\begin{align}
-\mathbb{E}_{H_0} \left[\am^\top\frac{\partial ( \psi_\tau(\tilde{r}(\xho))\hat{\sigma}_M(r(\xho)))}{\partial \xho}\right] = 
\left ( \begin{array}{c|c}
  E_{11} & E_{12} \\
  \hline
  E_{21} & E_{22}
 \end{array} \right )
\end{align}
  
then the matrix that we have to invert becomes
\begingroup\makeatletter\def\f@size{7.7}\check@mathfonts
\begin{align}
\left ( \begin{array}{c|c}
  E_{11} + 2\lambda\text{diag}(J''_K(\xho)_{1:t})) & E_{12} \\
  \hline
  E_{21} & E_{22} + 2\lambda 2K\mathbf{I}_{n-t}
 \end{array} \right )^{-1},
\end{align}
where $\mathbf{I}_{n-t}$ is the $(n-t)\times(n-t)$ identity matrix. 
The inverse of a block matrix can be written as 
\begin{align}
&\left ( \begin{array}{c|c}
  A & B \\
  \hline
  C & D
 \end{array} \right )^{-1} = \nonumber \\
 & \left ( \begin{array}{c|c}
  A^{-1}+A^{-1}B(D - CA^{-1}B)^{-1}CA^{-1} & -A^{-1}B(D - CA^{-1}B)^{-1} \\
  \hline
  (D - CA^{-1}B)^{-1}CA^{-1} & (D - CA^{-1}B)^{-1}
 \end{array} \right )
 \label{eq:blocks}
\end{align}
\endgroup
In our case,
\begin{align}
A & =  E_{11} + 2\lambda\text{diag}(J''_K(\xho)_{1:t}))), B = E_{12} \nonumber \\
C & = E_{21}, D = E_{22} + 2\lambda 2K\mathbf{I}_{n-t} \nonumber.
\end{align}

Let us call
\begin{align}
Z =&  D - CA^{-1}B = \nonumber \\
& E_{22} + 2\lambda 2K\mathbf{I}_{n-t} - \nonumber\\
& E_{21}(E_{11} + 2\lambda\text{diag}(J''_K(\xho)_{1:t})))^{-1}E_{12}.
\end{align}

We want to know what happens with $Z$ when $K\to\infty$. In the first place, we know that 
\begin{align}
\lim _{K\to\infty}J_K''(x) = 0 \text{ for } x\neq0
\end{align}
So
\begin{align}
Z =&  D - CA^{-1}B = \nonumber \\
& E_{22} + 4\lambda K\mathbf{I}_{n-k} -E_{21}E_{11}^{-1}E_{12}
\end{align}

Let us also call

\begin{align}
Y = E_{22} -E_{21}E_{11}^{-1}E_{12}.
\end{align}

If the eigenvalues of $Y$ are $\nu_1,\dots,\nu_n$, then the eigenvalues of $Z$ are $\nu_1 + 4\lambda K,\dots, \nu_n + 4\lambda K$.

$Z^{-1} = Q\Lambda^{-1}Q^{-1}$, so $Z^{-1}\to 0$ as $K\to\infty$.

$Z^{-1}$ appears in all the components of (\ref{eq:blocks})
\begin{align}
\left ( \begin{array}{c|c}
  A^{-1}+A^{-1}BZ^{-1}CA^{-1} & -A^{-1}BZ^{-1} \\
  \hline
  Z^{-1}CA^{-1} & Z^{-1}
 \end{array} \right ),
\end{align}

so in the end we have
\begin{align}
&\left ( \begin{array}{c|c}
  A & B \\
  \hline
  C & D
 \end{array} \right )^{-1} = 
&\left ( \begin{array}{c|c}
  A^{-1} & 0 \\
  \hline
  0 & 0
 \end{array} \right ) = &\left ( \begin{array}{c|c}
  E_{11}^{-1} & 0 \\
  \hline
  0 & 0
 \end{array} \right ).
 \end{align}
 
 From here, it is straight forward to arrive to (\ref{eq:ifL1}).
\end{proof}
\section*{Acknowledgment}

This work was supported by a SNF Grant: SNF-20FP-1\_151073 Inverse Problems regularized by Sparsity and by the project HANDiCAMS which acknowledges the financial support of the Future and Emerging Technologies (FET) programme within the Seventh Framework Programme for Research of the European Commission, under FET-Open grant number: 323944.

\ifCLASSOPTIONcaptionsoff
  \newpage
\fi

\bibliographystyle{IEEEbib}

\bibliography{bibliography}

\end{document}